\documentstyle[12pt,aaspp4]{article}

\begin{document}

\title{Young Globular Clusters and Dwarf Spheroidals}
\author{Sidney van den Bergh}
\affil{
Dominion Astrophysical Observatory \\
Herzberg Institute of Astrophysics \\
National Research Council of Canada \\ 
5071 West Saanich Road \\
Victoria, British Columbia, V8X 4M6 \\
Canada }

\begin{abstract}
Most of the globular clusters in the main body of the Galactic 
halo were formed almost simultaneously.  However, globular 
cluster formation in dwarf spheroidal galaxies appears to 
have extended over a significant fraction of a Hubble time.  
This suggests that the factors which suppressed late-time 
formation of globulars in the main body of the Galactic halo 
were not operative in dwarf spheroidal galaxies.  Possibly 
the presence of significant numbers of ``young'' globulars at 
R$_{\rm GC} > 15$~kpc can be accounted for by the assumption that 
many of these objects were formed in Sagittarius-like (but 
not Fornax-like) dwarf spheroidal galaxies, that were 
subsequently destroyed by Galactic tidal forces.  It would 
be of interest to search for low-luminosity remnants of 
parental dwarf spheroidals around the ``young'' globulars 
Eridanus, Palomar 1, 3, 14, and Terzan 7.  Furthermore 
multi-color photometry could be used to search for the 
remnants of the super-associations, within which outer 
halo globular clusters originally formed.  Such envelopes 
are expected to have been tidally stripped from globulars 
in the inner halo. 
\end{abstract}


\bigskip

\noindent \hspace{3in}\underline{The galaxy is, in fact, nothing but a} \newline
\smallskip
\hspace{3in}\underline{congeries of innumerable stars grouped} \newline
\smallskip
\hspace{3in}\underline{together in clusters.} \newline
\smallskip
\hspace{3in}Galileo (1610)

\section{Introduction}
	The vast majority of Galactic globular clusters 
appear to have formed at about the same time (e.g. Richer 
et al. 1996, Sarajedini, Chaboyer \& Demarque 1997, Stetson, 
VandenBerg, \& Bolte 1996).  For various caveats that apply 
to the age determinations of globular clusters the reader is 
referred to VandenBerg (1999).  Even the outer halo globular 
cluster NGC 2419 (Harris et al. 1997), located at a Galactocentric 
distance of  $\sim 0.10$~Mpc, and the majority of the globulars in 
the Fornax dwarf at R$_{\rm GC} = 0.14$~Mpc (Buonanno et al. 1998), have 
approximately the same age as the bulk of the globulars in the 
main body of the Galactic halo.  Similar ages are also found for 
the globular clusters associated with the Large Magellanic Cloud 
(Olsen et al. 1998, Johnson et al. 1998). 

	However, a few globular clusters appear to have significantly 
smaller ages.  Presently known (or suspected) ``young'' globulars are:  
Palomar 12 (Stetson et al. 1989, Rosenberg et al. 1998b), Ruprecht 
106 (Buonanno et  al. 1993), IC 4499 (Ferraro et al. 1995), Rup 106, 
Arp 2, Pal 12 and Terzan 7 (Richer et al. 1996), Pal 12, Terzan 7, 
Rup 106, Arp 2 (Fusi Pecci et al. 1995), Arp 2 and Terzan 7 
(Montegriffo et al. 1998), Pal 1, Pal 3, Pal 4, and Eridanus 
(Stetson et  al. 1999), Fornax No. 4 (Marconi et al. 1999), 
Palomar 14 (Sarajedini 1997), and perhaps NGC 4590 (= M 68).  
The latter cluster was regarded as ``young'' by Chaboyer et al. 
(1996), but was considered to be of average age by Richer et  
al. (1996).  Table 1 gives a compilation of data on the globular 
clusters that have been listed as being young.  It should be emphasized 
that the calculated age differences between ``young'' clusters, and more 
typical Galactic globular clusters, might be reduced if [$\alpha$/Fe] 
is smaller than was assumed. 
Available data suggest (Sarajedini 1999) that the age 
range of globular clusters increases with metallicity from perhaps 1.5 
Gyr to 2 Gyr at [Fe/H] $\sim -1.6$, to 2--3 Gyr at [Fe/H] $\sim -1.0$.

	It is presently not clear if the cluster Pal 1 (Rosenberg et 
al. 1998a, c) at R$_{\rm GC} = 11.7$~kpc, Z = +3.7 kpc, with [Fe/H] = -0.8, 
M$_{\rm V} =  -2.5$, and an age of  $\sim 7$ Gyr is, 
in fact, an open or a globular 
cluster.  Perhaps it represents a transitional type of object.  
Because of its small radius and low population the evaporation 
time-scale for Pal 1 is only  $\sim 0.7$ Gyr (Rosenberg et al. 1998a).  
A large initial population of such low-mass clusters might have become 
extinct.  Another object that could be intermediate between open and 
globular clusters is Lyng\aa\ 7 (Ortolani, Bica \& Barbuy 1993, Tavarez \& 
Friel 1995).  The observation that the very luminous outer halo globular 
cluster NGC 2419 is old (Harris et al. 1997), whereas younger outer 
halo globulars, such as Pal 3, Pal 4 and Eridanus, are relatively 
young (Stetson et al. 1999), suggests the possibility that the mean 
luminosity (mass) with which globular clusters formed in the outer 
halo might have decreased with time, from values that are characteristic 
of typical globulars, to values that are more similar to those of 
the typical open clusters which are still  being formed in the Galactic 
disk at the present time.  A possible argument against this idea is 
that there seems to be no obvious correlation between the ages and 
the luminosities of the globular clusters in Sagittarius (Montegriffo 
et al. 1998) and Fornax (Marconi et al. 1999).  Such a correlation 
might have been expected if the globulars in the outer halo had been 
formed in dwarf spheroidals which were subsequently disrupted.  
It is presently not entirely clear why there was a guillotine-like 
cut-off in the rate of globular cluster formation in the inner 
Galactic halo, while observations of Ter 7 suggest that such cluster 
formation continued for up to  $\sim 7$ Gyr in dwarf spheroidals.  
Possibly Searle-Zinn (1978) fragments, with orbits that took them to 
R$_{\rm GC} < 15$~kpc, were tidally destroyed or stripped of gas on a 
relatively short time-scale.  Some of these ideas have previously 
been discussed by Freeman (1990) and by Bassino, Muzzio \& Rabolli (1994).

	A listing of data on presently known ``young'' globular 
clusters, based mainly on Harris (1996), which was supplemented 
by data listed 1999 June 22 at 
http://physun.physics.mcmaster.ca/Globular.html, is given in Table 1.  
With the exception of NGC 4590 at R$_{\rm GC} = 10.0$~kpc 
(which is not ``young'' according to Richer et al.) all of these 
globulars are located in the 
outer halo at R$_{\rm GC} > 15.0$~kpc.  Many of these clusters are seen to 
have below-average luminosities (van den Bergh 1998).  Eleven of 
the 40 globulars situated at R$_{\rm GC} > 15.0$~kpc are presently known to 
be ``young''.  The true fraction of such objects of below-average age 
in the outer halo is probably even greater.  This is so because 
high-quality color-magnitude diagrams, that reach down below the 
main sequence turnoff, are not yet available for quite a few of the 
distant clusters in the outer halo of the Milky Way system.

\section{Globulars in Dwarf Spheroidals}
	From the updated compilation of Harris (1996) it is found 
that the Galactic halo contains 31 globular clusters with 
R$_{\rm GC} > 15.0$~kpc.  Furthermore there are four globulars associated 
with the Sagittarius dwarf spheroidal galaxy (Ibata, Gilmore  \& 
Irwin 1994) at R$_{\rm GC} = 18.6$.  A fifth (Palomar 12) has both a velocity, 
and a position on the sky, which suggests that it was originally 
associated with, and subsequently stripped from, the Sagittarius 
dwarf (Irwin 1999).  An additional five globular clusters are 
situated in the Fornax dwarf spheroidal galaxy at R$_{\rm GC} = 0.14$~Mpc.  
The total cluster population (excluding globulars associated 
with the LMC and SMC) beyond 15 kpc is therefore 40.  Table 1 
shows that 11 of these objects, i.e. a quarter of the total, are 
now thought to be ``young''.  Within Fornax one (Fornax No. 4) out 
of five clusters is ``young'' (Marconi et al. 1999).  In Sagittarius 
two out of four (or three out of five if Pal 12 is included) are 
``young''. These numbers are consistent with the notion that \underline{all 
globular clusters, that are presently located in the outer halo 
of the Galaxy, might} \underline{originally have formed in dwarf spheroidals} 
(most of which subsequently suffered destruction by Galactic tides).  
Alternatively Lee \& Richer (1992) proposed that young clusters, 
such as Pal 12 and Rup 106, might have been tidally captured from 
the Magellanic Clouds.  However, the metallicity of these clusters 
appears too low to be consistent with this hypothesis.  Pal. 12 
has [Fe/H] = -1.0 and Rup. 106 has [Fe/H] = -1.45 (Brown, Wallerstein 
\& Zucker 1997).  Both of these values are lower than those presently 
prevailing in the Clouds of Magellan.  They would therefore have to 
have been formed long ago, before the LMC and SMC were enriched 
significantly in heavy elements.  However, (admittedly uncertain) 
orbital simulations by Byrd et al. (1994) suggest that the Magellanic 
Clouds were still located near M 31  $\sim 10$ Gyr ago, and were not 
captured by the Galaxy until  $\sim 6$ Gyr ago.  Furthermore, any physical 
association between Pal 12 and the Magellanic Clouds would conflict 
with Irwin's (1999) suggestion that this object was, in fact, stripped 
from the Sagittarius dwarf spheroidal galaxy.  The Sagittarius dwarf 
might originally have been a Searle-Zinn fragment (Searle \& Zinn 1978), 
which formed in the outer Galactic halo, and that was subsequently 
scattered into a shorter period orbit by gravitational interaction with 
the Magellanic Clouds (Zhao 1998, van den Bergh 1998).  However, an 
argument against this hypothesis (Jiang \& Binney 1999) is that the 
velocity of encounter between the Clouds and the Sagittarius dwarf 
may have been too large for the Magellanic Clouds to deflect it through
 a significant angle.

	Inspection of Table 1 shows that three out of 11 (27\%), or four 
out of 11  (36\%) if Pal 12 is included, of all ``young'' globulars are 
associated with known dwarf spheroidal galaxies.  This suggests that it 
might be worthwhile to use the digitized version of the \underline{Second 
Palomar Sky Survey} to search for additional 
very faint (and so far undiscovered) dwarf 
spheroidals surrounding the other ``young'' globulars listed in Table 1.  
Such a search would, however, be quite difficult because the postulated 
faint spheroidals associated with ``young'' globulars are relatively nearby, 
and are therefore expected to subtend a large angle on the sky. In view of 
this problem it might turn out to be more efficient to use multi-color 
photometry of large fields surrounding each ``young'' globular to search 
for these hypothetical parental objects.

	Ambartsumian (1955) wrote ``It seems probable that the development 
of an association involves both expansion and the formation of one or more 
open clusters.''  He cited the example of the cluster IC 348 in the 
association Per OB2. Perhaps the best known case of clusters in an 
association is the (positive energy) association Per OB1 within which 
are located the two (negative energy) clusters h Per (NGC 869) and 
$\chi$ Per (NGC 884).  [For a complete listing of clusters associated with 
associations the reader is referred to the compilation by Ruprecht (1966).]  
The existence of stable clusters within associations suggests that 
\underline{globular clusters may, at the time of their formation, 
also have been embedded in massive} \underline{associations}.  
Parts of such associations might have 
survived at large Galactocentric distances.  However, most of the 
associations surrounding globulars in the inner part of the Milky Way 
system would probably have been dispersed long ago by Galactic tidal 
forces.  It should be noted that such globulars in the main body of the 
halo are expected to be surrounded by tidal debris consisting of material 
detached from the clusters themselves (Grillmair 1998). From an observational 
point of view it will be difficult to distinguish between the remnant 
of a primordial association and tidally stripped material of similar age 
and chemical composition.  Since tidal forces in the Magellanic Clouds 
will generally be lower than they are in the Galaxy, it might be easier 
to find ancestral super-associations around the outer globular clusters 
of the LMC, than it would be to find traces of such structures near Galactic 
globulars.  Binary and multiple clusters [see Pietrzy\'{n}ski and Udalski 
(1999) for a review] in the SMC might be tracers of such large old 
associations.  Bica et al. (1998) have observed the unique old Large 
Cloud cluster ESC 121-SC02, which has an age of  $\sim 9$ Gyr, to be embedded 
in a field population of similar age.  Since ESO 121-SC02 is located in 
the outer reaches of the LMC it is possible that these stars are the 
remnants of the association within which this cluster formed.  In this 
connection it is of interest to note (Schweizer 1999) that the young knot 
S, in the merging galaxy NGC 4038, has a cluster-like core with 
M$_{\rm V} = -16$, 
that is embedded in a power-law like envelope containing hundreds of 
stars out to a distance  $\sim 450$~pc.  This may represent an example of a 
``young globular cluster'' that is still embedded in its ancestral super 
association.

\section{Globular Cluster Radii}
Figure 1 shows a plot (based on an update of the data in Harris 1996) of 
the half-light radii R$_{\rm h}$ of globular clusters as a function of their 
Galactocentric distances R$_{\rm GC}$.  For isolated clusters such half-light 
radii are valuable diagnostic tools because they remain almost independent 
of cluster evolution over  $\sim 10$ cluster relaxation times (Spitzer \& 
Thuan 1972, H\'{e}non 1973, Lightman \& Shapiro 1978, 
Murphy, Cohn \& Hut 1990).  
The data plotted in Figure 1 show that globular clusters situated in 
the outer halo are systematically larger than those located closer to 
the Galactic center.  That this relationship is not entirely due to the 
tidal destruction of large globular clusters at small values of 
R$_{\rm GC}$ is demonstrated by the fact that no small 
compact clusters occur at large values of R$_{\rm GC}$, 
even though such 
\noindent compact objects would easily have survived 
the weak tidal field in the outer halo.  Van den Bergh (1995) showed that 
an even tighter relationship exists between the sizes of Galactic globulars 
and their perigalactic distances.  A similar relationship between half-light  
radii and galactocentric distances is observed for both open clusters, and 
globular clusters, in the Large Magellanic Cloud (LMC) (van den Bergh 1994).

Inspection of Figure 1 shows that the distribution of ``young'' globular 
clusters, and of the globular clusters associated with the Sagittarius 
system, does not differ from that of Galactic globulars.  However, the 
globular clusters in the Fornax dwarf are located well to the right of the 
distribution for globulars associated with the Galaxy.  This result suggests 
(not surprisingly) that \underline{the Fornax system, 
at a distance of 0.14 Mpc, developed independently, and was never} 
\underline{part of the Milky Way protogalaxy}.  
Figure 2, which is based on the data in Table 2 of 
van den Bergh (1984), shows that the globular clusters in the Large 
Magellanic Cloud exhibit a close relationship between galactocentric 
distance and half-light radius R$_{\rm h}$ .  This shows that the LMC cluster 
system developed independently from the Galactic globular cluster system.  
Intercomparison of Figure 1 and Figure 2 shows that the Large Cloud 
globulars are, at a given galactocentric distance, systematically larger 
than most of their Galactic counterparts.

\section{Chemical Signature of Slow Evolution}
 	Brown, Wallerstein \& Zucker (1997) have made spectroscopic 
abundance determinations for a few stars in the ``young'' globular 
clusters Rup 106 and Pal 12.  They find that the abundance ratios in 
these clusters are peculiar.  Brown et al.  note that the ratios of 
$\alpha$-elements [defined as the even-even nuclei from Mg to Ti] to iron 
are not enhanced over their solar ratios, as they are in most globular 
clusters and metal-poor halo stars.  The most straightforward explanation 
for this is that the stars in Rup 106 and Pal 12 were formed from gas 
which had been enriched in iron on a slow time-scale by supernovae of 
Type Ia.  For Rup 106 Brown et al. find $[{\rm Fe/H}] = -1.45 \pm 0.10$ and 
$[{\rm O/Fe}] = 0.0 \pm 0.1$.  Only a small number of high-velocity stars are 
presently known to exhibit such a low [$\alpha$/Fe] abundance.  Nissen \& 
Schuster (1997) find that the smallest values of 
[$\alpha$/Fe] occur for those halo 
stars that have the largest values of R(apo) and the greatest distances 
from the Galactic plane.  However, this conclusion is not confirmed by 
Stephens (1999).  The reason for this disagreement is not yet understood, 
but might be related to sample selection.  The stars in the sample of 
Nissen \& Schuster were biased towards objects with low orbital velocities, 
and therefore have ${\rm R(peri)} < 1.0$~kpc, 
whereas the stars in Stephens' sample 
were selected to have large R(apo).  All but one of the members of the 
Stephens' sample have ${\rm R(apo)} > 25$~kpc.  
Gilmore \& Wise (1998) point out 
that tidal capture of a low-density dwarf galaxy will result in its 
disruption. As a result it is highly unlikely that stars from such a dwarf 
will sink to very small R(peri) distances.

	Both theory (Leonard \& Duncan 1990) and observation (Blaauw 1961, 
Gies \& Bolton 1986) lead one to expect that binaries will be rare (or absent) 
among stars that have undergone a significant amount of dynamical 
acceleration. \underline{Binary/multiple stars that 
have very} \underline{high space velocities 
are therefore likely to be true halo stars that were formed during the first} 
\underline{violent (most chaotic) phase of Galactic evolution}.  
One of the few presently 
known examples of such a high velocity binaries is HD 134439/40 (King 1997).  
The (retrograde) space velocity of these objects is 565~km~s$^{-1}$, and their 
metallicity is [Fe/H]  $\sim -1.5$.  
King notes that [Mg/Fe], [Si/Fe], and [Ca/Fe] 
in HD 134439/40 are consistently  $\sim 0.3$ 
dex lower than they are in the vast 
majority of metal-poor stars in the Galactic halo. According to Carney et al. 
(1994) HD 134439/40 have plunging orbits with ${\rm R(apo)} = 43$~kpc and 
${\rm R(peri)} = 4$~kpc, 
which suggests that they were formed in the outer halo, or that they 
were captured rather late in the evolutionary history of the protoGalaxy.

Another (rare) example of an $\alpha$-deficient 
high velocity star is the subgiant BD +80$^\circ$~245, 
which has [$\alpha$/Fe]$ = -0.29 \pm 0.02$ (Carney et al. 1997).  According 
to Carney et al. this object is on a plunging orbit with 
${\rm R(apo)} = 22$~kpc.  Other stars with low 
[$\alpha$/Fe] are HD 6755 and HD 108577 (Carney 1999).  It 
should, however, be emphasized that not all $\alpha$-poor 
high-velocity stars are 
halo objects.  King (1997) has, for example, drawn attention to the star 
BD +3$^\circ$~740, with [Fe/H] $\sim -3$, in which [Mg/Fe] and 
[O/Fe] are  $\sim 0.5$ dex 
lower than in most metal-poor field stars (Fuhrmann, Axer \& Gehren 1995).  
This object, which has ${\rm R(apo)} = 10$~kpc 
and ${\rm R(peri)} = 2$~kpc (Carney et al. 1994), 
is clearly not a member of the outer halo population of the Galaxy.  
The fact that the $\alpha$-elements in the 
aforementioned objects are only about 
half as abundant, relative to iron, as they are in the overwhelming majority 
of metal-poor stars indicates that they were formed in an environment in 
which fast enrichment of $\alpha$-elements by SNe~II 
was relatively less important 
than the slower enrichment of Fe produced by SNe of Type Ia.  Alternatively, 
one can make the \underline{ad} \underline{hoc} 
assumption that the environment in which these stars 
were produced did not favor the formation of the massive stars that are the 
progenitors of $\alpha$-element producing supernovae of Type II.  
Browne et al. suggest that the apparent deficiency of 
s-process elements that is observed 
in Rup 106 might be due to the fact that such ``young'' globulars formed 
after SNe~Ia had contributed most of their iron, but before the s-process 
contributors had evolved and shed their outer layers.

	It is noted in passing that the value [Fe/H] $\sim -1.5$ 
that King (1997) obtain for HD 134439/40 is 
indistinguishable from $\langle[{\rm Fe/H}]\rangle = -1.49 \pm 0.11$ 
that Rodgers \& Paltoglou (1984) found for the seven Galactic globulars 
that are known to be in retrograde motion.  The old globular cluster NGC 
3201, which is in a retrograde orbit and has 
$[{\rm Fe/H}] = -1.42 \pm 0.3$, 
exhibits the usual excess of $\alpha$-elements 
(Gonzalez \& Wallerstein 1998).  
This observation supports the hypothesis that 
[$\alpha$/Fe] for metal-poor objects 
is mostly determined by their ages, rather than by their orbital 
characteristics.  It is, however, puzzling (Wallerstein, Brown, \& 
Gonzalez 1998) that two stars in M 54 (= NGC 6715), which is located in 
the Sagittarius dwarf, appear to be slightly oxygen deficient having [O/Fe] 
values of $-0.23 \pm 0.16$ and $-0.10 \pm 0.17$, respectively.  This is so 
even though
M 54 appears to be approximately the same age 
($\Delta T = +1.0 \pm 2.0$~Gyr) 
as the old globular cluster Ter 8.  Taken at face value these results 
might tend to indicate that the apparent oxygen deficiency in M 54 was 
due to a stellar luminosity function deficient in high-mass SN~II 
progenitors, rather than to an above-average contribution of Fe by 
SNe Ia.

\section{Other Evidence for Young Stars in the Galactic Halo}

Perhaps the most unambiguous evidence for the existence of massive 
stars in the Galactic halo is provided by the observation (van den 
Bergh \& Kamper 1977) that \underline{the progenitor of Kepler's supernova of 
1604 had a very high space velocity}.  Adopting a conservative 
distance of 4.5~kpc Bandiera \& van den Bergh  (1991) derived a 
space velocity of $278 \pm 12$~km~s$^{-1}$ for this remnant.  Furthermore 
Bandiera (1987) showed that the observed velocities in, and the 
spatial distribution of, the nebulosity in the remnant of SN 1604 
exhibit a pattern that is best reproduced by the stellar wind of a 
massive evolving star, that is interacting with the low-density 
interstellar medium in its environment.  Bandiera concludes that 
the progenitor of Kepler's supernova was a massive object 
(${\rm M} > 10\ {\rm M}_{\sun}$) that is currently losing mass at a rate of  
$\sim 5 \times 10^{-5}\ {\rm M}_{\sun}$~yr$^{-1}$.  
He shows that this object must have left 
the Galactic plane with a velocity of  $\sim 340$~km~s$^{-1}$ some 
$3 \times 10^6$ yr ago.  The remnant of Kepler's supernova is presently 
situated at ${\rm Z} = +530$~pc.  
\underline{The existence of runaway OB stars, 
and of Kepler's supernova, demonstrate that not all} 
\underline{high-velocity 
halo objects belong to an old stellar population.}

	The fact that the outermost Galactic halo has a (small) 
retrograde velocity, in conjunction with the observation of 
streaming motions in the halo (Majewski, Hawley \& Munn 1996, 
Majewski, Munn \& Hawley 1996), suggests that a large fraction 
of the outermost halo population was accreted.  Presumably 
the rate at which this accretion took place was initially 
quite high.  However, some of this capture by the protoGalaxy 
appears to have taken place only 3--10 Gyr ago.  From a survey 
of young blue halo stars Unavane, Wyse \& Gilmore (1996) 
conclude that up to 10\% of the Galactic halo may consist of 
accreted stars.  Preston, Beers \& Shectman (1994) have reviewed
 evidence for the presence of such young and intermediate-age 
stars in the Galactic halo.  They conclude that blue metal-poor 
stars in the halo are the bluest main sequence members of a 
metal-poor intermediate-age population, that was probably 
captured in the form of dwarf spheroidal galaxies.  They 
estimate that 4--10\% of the Galactic halo consists of such 
intermediate-age accreted stars.  The Galactic halo has 
M$_{\rm V} = -18.4$ (Suntzeff 1992).  The total luminosity of the 
captured material therefore has  $-15.9 < {\rm M}_{\rm V} < -14.9$.  
For comparison it is noted that Sagittarius presently has a 
(very uncertain) luminosity of M$_{\rm V} = -13.8$, i.e. the total 
amount of captured stellar material in the halo is probably 
equivalent to 3--7 Sagittarius dwarfs.  (A smaller number of 
``Sagittarius captives'' would be obtained if tidal stripping 
has reduced the original luminosity of Sagittarius).  Since
 4--5 globulars appear to be associated with Sagittarius, the 
corresponding number of captured globulars would be 12--35.  
This number is comparable to the total number of Galactic 
globulars with R$_{\rm GC} > 15.0$.  
This shows that \underline{the entire population 
of globulars in the outer halo (or at least a significant fraction 
of it)} \underline{might have been formed in dwarf 
spheroidals and subsequently 
accreted by the Galaxy}.  A possible argument against this view is 
provided by the fact that very large outer halo clusters such as 
Pal 4 (R$_{\rm h} = 16$~pc), Pal 3 (R$_{\rm h} = 17$~pc) and 
Pal 14 (R$_{\rm h} = 24$~pc) 
might not have been able to survive the tidal stresses within dwarf 
spheroidal galaxies.  Interactions between globulars within 
spheroidals (Rogers \& Roberts 1994) might also reduce the sizes 
of such clusters.  Shetrone, Bolton \& Stetson (1998) find that 
some red giants in the Draco dwarf spheroidal have significantly 
lower values of [Ba/Fe] than do field halo stars of similar 
[Fe/H].  This indicates that the dominant stellar population 
in the Galactic halo was probably not derived from disintegrated 
dwarf spheroidals.

\section{Globulars as Nuclei of Dwarf Spheroidals}
	It has been argued (Larson 1996) that M 54 might be the nucleus 
of the Sagittarius system.  However, van den Bergh (1986) has 
shown that the fraction of spheroidal galaxies that have nuclei 
is strongly dependent on luminosity, and ranges from 100\% at 
M$_{\rm V} = -17$, to  $\sim 15$\% at M$_{\rm V} = -12$.  
Adopting M$_{\rm V} = -13.8$ for 
Sagittarius, and assuming the same nuclear frequency as is 
found for spheroidal galaxies in the Virgo cluster, one finds 
that the a priori probability that Sagittarius had a nucleus 
is slightly less than 50\%.  A higher probability would, however, 
be obtained if the luminosity of Sagittarius has, because 
of tidal stripping, decreased over time.  Very recently Dinescu, 
Girard \& van Altena (1999) have noted that the orbital 
characteristics of  $\omega$ Centauri are consistent with the hypothesis 
that this object was also once the nucleus of a dwarf galaxy 
that was subsequently destroyed by tidal forces.  If this 
hypothesis is correct, then two of the most luminous Galactic 
globulars might both have started their existence as the nuclei 
of dwarf galaxies.  This suggests the possibility that all 
globular clusters started out as the cores of more extended 
stellar distributions.  In this respect they might have resembled 
OB associations, which sometimes have young open clusters 
in their cores.

\section{Associations and Giant Clusters}
	The 30 Doradus nebula, and its central ionizing cluster 
R 136, is the nearest (and best-studied) super star forming region.  
From UBV photometry Malamuth \& Heap (1994) derived a total 
mass of 16\,800 M$_{\sun}$ for R 136.  Neglecting mass loss by evolving 
stars, and assuming M/L$_{\rm V}  = 3$ for clusters with 
ages $\sim 10$ Gyr, 
yields M$_{\rm V} = -4.5$ for R 136 at age 10 Gyr.  This is 2.86 mag 
lower than the value $\langle{\rm M}_{\rm V}\rangle = -7.36$, 
that Harris (1991) finds for 
all Galactic globulars.  The mass gap between R 136 and typical 
globulars would have been even greater if the mass loss by 
evolving stars had been taken into account.  In other words, 
R 136 will (if it survives for 10 Gyr) be more than an order 
of magnitude less luminous (massive) than typical Galactic 
globulars.  This suggests that the formation of a typical 
Galactic globular was probably accompanied by a much greater 
burst of star formation than that which is presently 
observed in the 30 Doradus region.

	The structure of the 30 Doradus region is complex 
(Walborn 1991).  Rubio et al. (1998) show that the central 
dense cluster R 136 has triggered a second generation of 
star formation.  These authors speculate that 30 Dor will 
eventually become a giant H II shell resembling N 11 in 
the LMC and NGC 604 in M 33.  This suggests that the even 
greater bursts of star formation, that produce typical 
globular clusters, will give rise to a huge ``super association''.  
Presumably such an association of secondary stars will, 
over a Hubble time, be tidally stripped from globulars in the 
main body of the Galactic halo.  However, such associations 
might have survived around some of the globular clusters in 
the outer halo of the Galaxy where tidal forces are weak.  
This is so because it requires many perigalactic passages to 
remove the bulk of the stars beyond the tidal radius.  Over 
the lifetime of the cluster there will be a continual flow of 
new stars out of the cluster envelope into the extra-tidal 
region.  Combined with the long orbital periods of objects in 
the outer halo this will substantially increase our chances 
of detecting these stars. A multi-color search for the envelopes 
surrounding remote globulars might profitably be undertaken 
with the large area detectors that have recently become available.  
It should, perhaps, be emphasized that most of the stars in 
dwarf spheroidal galaxies were probably not produced during 
secondary bursts of star formation that were triggered by young 
globular clusters.  This is so because most dwarf spheroidals 
do not have globular clusters embedded within them. Furthermore 
the globulars that are associated with the Fornax and Sagittarius 
systems contribute only 2.7\% and 3.0\%, respectively, to the 
integrated luminosities of their parent galaxies.  (Their 
fractional contribution to the total mass is probably even 
smaller).  A physical distinction between super associations 
triggered by formation of a young globular, and dwarf spheroidals 
is that associations do not contain dark matter, whereas (most?) 
dwarf spheroidals do (Moore 1996).

\section{Conclusions}
\begin{enumerate}
\item	Many of the Galactic globular clusters with 
R$_{\rm GC} > 15$~kpc might originally have 
formed in dwarf spheroidal galaxies.  
Formation of globular clusters appears to have continued for 
up to 7 Gyr in dSph galaxies and in the Galactic halo beyond 
R$_{\rm GC} = 15$~kpc.  This contrasts with the situation in the main 
body of the Galactic halo in which globulars have only a small 
age range.  It is not clear why the guillotine that cut off 
cluster formation at R$_{\rm GC} < 15$~kpc did not operate in the outer 
halo and in dwarf spheroidal galaxies.

\item	In the outer halo there may be some evidence for a 
decline of the mean luminosity of new born globular clusters 
with age.  A few late forming objects, like Pal 1, have 
characteristics that make them appear to be intermediate 
between open clusters and globular clusters.

\item	The luminous globular clusters $\omega$ Centauri and M 54 
(= NGC 6715) may once have been the nuclei of dwarf spheroidal 
galaxies.  Since about a quarter of all presently known 
``young'' globular clusters are located in dwarf spheroidal 
galaxies, it might be worthwhile to search for the remnants 
of faint dSph galaxies surrounding the ``young'' globulars 
in the outer halo.

\item	Negative energy (stable) clusters are situated in the 
cores of positive energy expanding Galactic associations.  
Furthermore R 136 appears to have induced secondary star 
formation in the 30 Doradus region.  This suggests that an 
envelope of secondary stars might have been formed around 
globular clusters. Galactic tidal forces will have removed 
such shells from globular clusters at small and intermediate 
galactocentric distances.  However, it might still be possible 
to find remnants of shells of secondary star formation around s
ome globulars with R$_{\rm GC} > 15$~kpc.

\item	Available data on the half-light radii of globular clusters 
suggest that the LMC and the Fornax dwarf formed separately from 
the Milky Way, but that the Sagittarius dwarf may have formed 
as part of the outer protoGalaxy.

\item	Stars with large space velocities need not all be old.  
Kepler's supernova is the prototype of high-velocity halo stars 
that were dynamically ejected into the Galactic halo.  Only very 
high velocity binaries like HD 134439/40 or outer halo clusters, 
such as Rup 106 and Pal 12, are likely to provide insight into 
the chemical evolution of the outer Galactic halo.

\end{enumerate}

\begin{acknowledgements}
	It is a pleasure to thank many of my colleagues for 
interesting discussions and e-mail regarding some of the problems 
discussed above.  I am also indebted to the referee (Carl Grillmair) 
for a number of helpful suggestions.
\end{acknowledgements}

\begin{deluxetable}{lrrrrc}
\tablecaption{PROPERTIES OF ``YOUNG'' GLOBULAR CLUSTERS\tablenotemark{a}}
\tablehead{
\colhead{Name} &
\colhead{R$_{\rm CG}$} &
\colhead{M$_{\rm V}$} &
\colhead{[Fe/H]} &
\colhead{R$_{\rm h}$} &
\colhead{dSph} \\
\colhead{} &
\colhead{(kpc)} &
\colhead{} &
\colhead{(dex)} &
\colhead{(pc)} &
\colhead{} }
\startdata
Arp 2       &  29 &-5.3 &-1.76 & 15.9 & Sgr \nl

Eridanus    &  90 &-5.1 &-1.46 & 10.5 & ... \nl  

Fornax No. 4& 138 &-7.4 &-1.35 &  3.0 & For \nl

IC 4499     &  19 &-7.3 &-1.60 &  8.2 & ... \nl

NGC 4590\tablenotemark{b}   &  10 &-7.3 &-2.06 &  4.5 & ... \nl

Palomar 1\tablenotemark{c}  &  11 &-2.5 &-0.60 &  2.2 & ... \nl

Palomar 3   &  93 &-5.7 &-1.66 & 17.8 & ... \nl

Palomar 4   & 109 &-6.0 &-1.48 & 17.2 & ... \nl

Palomar 12  &  19 &-4.5 &-0.94 &  7.1 & Sgr ? \nl

Palomar 14  &  74 &-4.7 &-1.52 & 24.7 & ... \nl

Ruprecht 106&  21 &-6.4 &-1.67 &  6.8 & ... \nl

Terzan 7    &  23 &-5.0 &-0.58 &  6.5 & Sgr \nl
\enddata
\tablenotetext{a}{Data mainly from 1999 update of Harris (1996).}
\tablenotetext{b}{$=$ M 68.  According to Richer et al. 
(1996) this cluster is not ``young''.}
\tablenotetext{c}{Old open cluster?  Data from Rosenberg et al. (1998c).}
\end{deluxetable}

\begin{figure}
\plotone{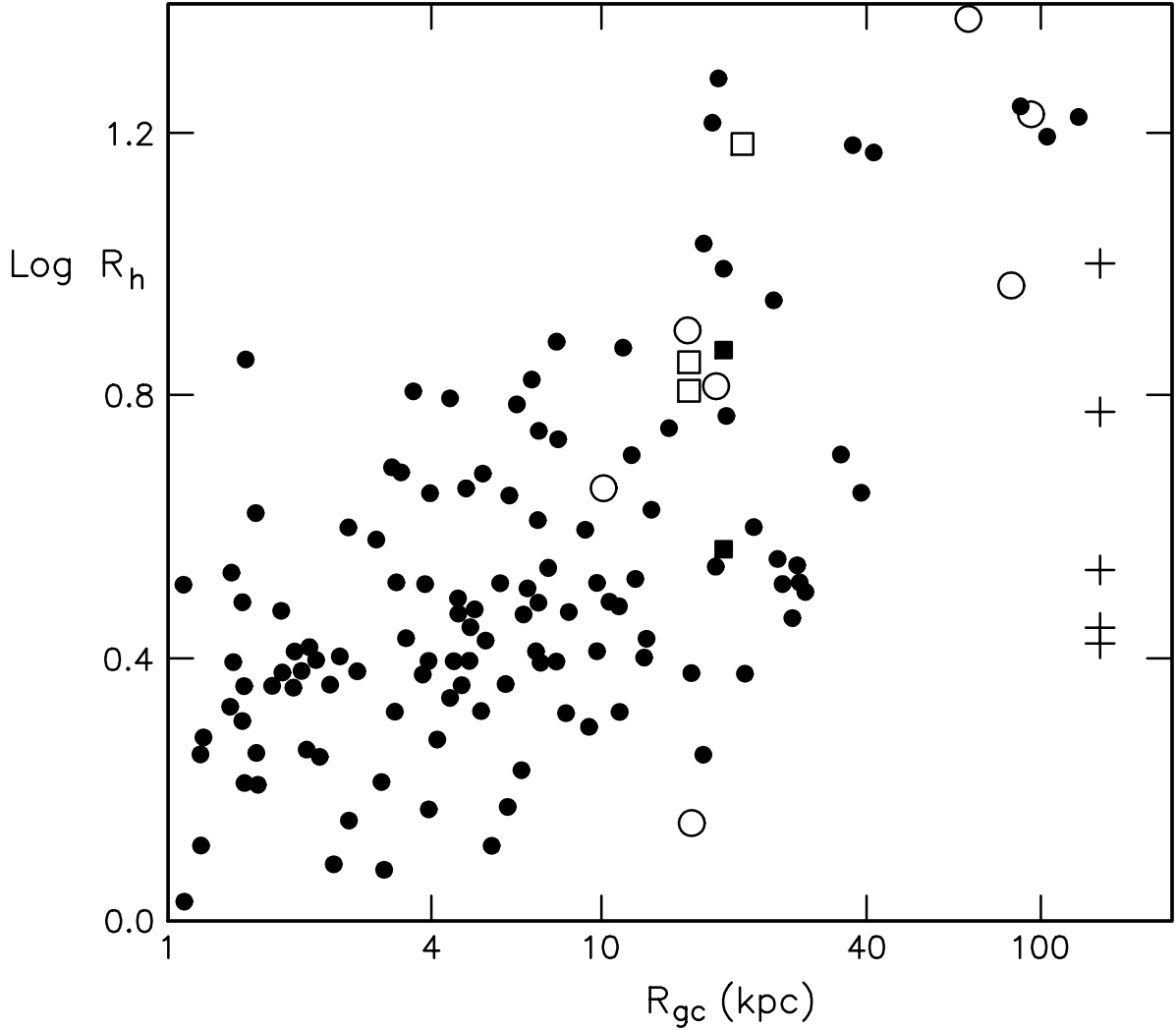}
\caption{Cluster half-light radii as a function of Galactocentric distance.
The Figure shows that the half-light radii of Galactic globular clusters 
are strongly correlated with their Galactocentric distances.  No 
difference is seen between the positions of ``young'' globulars 
(open circles) and globulars associated with the Sagittarius dwarf 
(open squares), and those associated with the Galaxy (filled circles).  
However, clusters associated with the Fornax system (plus signs) and 
the LMC (not shown) have a different distribution.  This suggests that 
the Galaxy, the LMC and Fornax were already distinct stellar systems 
at the time when they started to form their globular clusters.}
\end{figure}

\begin{figure}
\centerline{\epsfysize=5in%
\epsffile{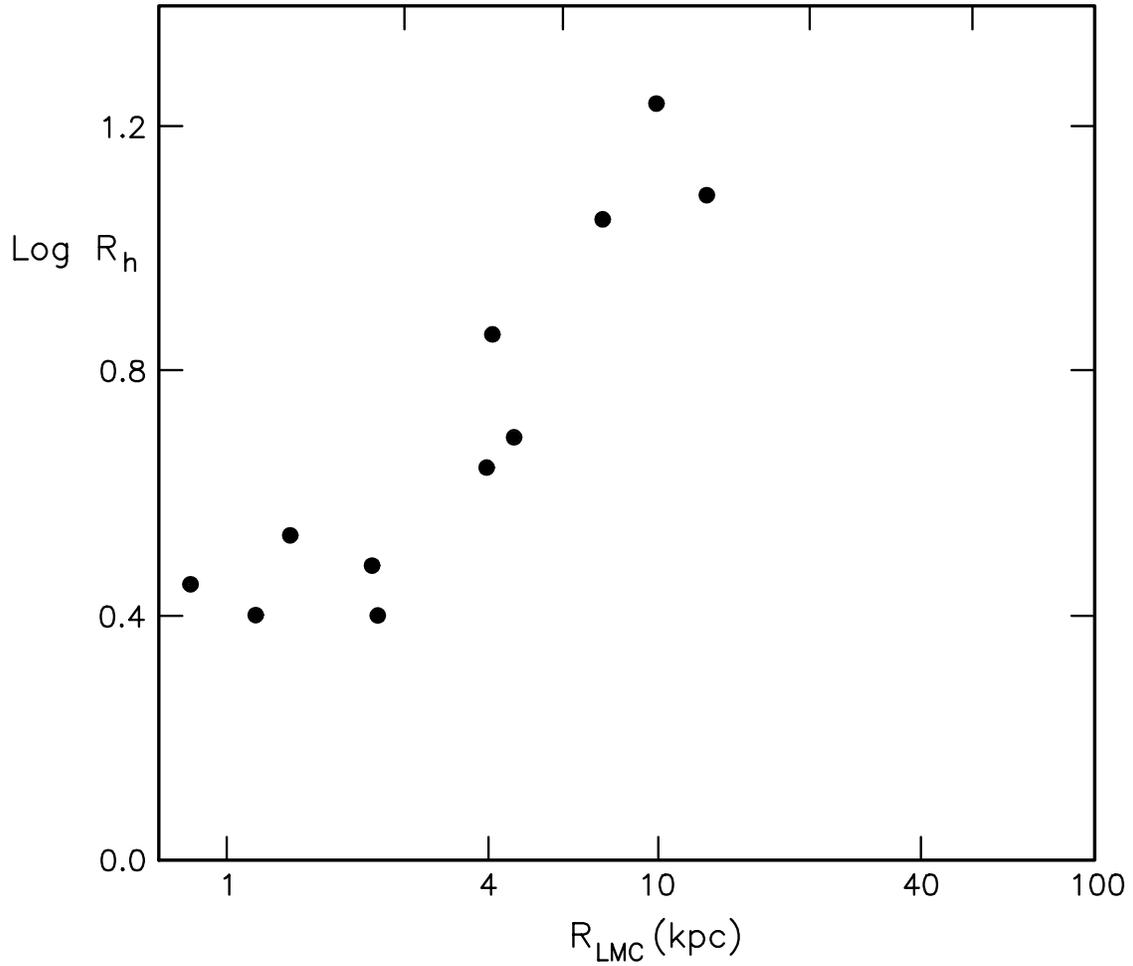}}
\caption{Half-light radii versus galactocentric distances for globular 
clusters associated with the Large Magellanic Cloud.  The figure shows 
that R$_{\rm h}$ increases with distance from the center of the LMC.  
This shows that the Large Cloud cluster system formed independently 
from the Galactic globular cluster system.  At a given galactocentric 
distance the Large Cloud globulars tend to be larger than their Galactic 
counterparts.}
\label{fig2}
\end{figure}

\end{document}